\documentclass[nofootinbib,aps,prc,superscriptaddress,showpacs,amssymb,amsmath,amsfonts,floatfix]{revtex4-1}
\usepackage{graphicx}
\usepackage{xcolor}

\begin{document}

\title{High Energy Behaviour of Light Meson Photoproduction}

\newcommand*{\ODU}{Department of Physics, Old Dominion University, 
	Norfolk, VA 23529, USA}
\newcommand*{\PNPI}{ Petersburg Nuclear Physics Institute, NRC Kurchatov Institute, Gatchina, 
    St.~Petersburg, 188300, Russia}
\newcommand*{\GWU}{Institute for Nuclear Studies, Department of Physics,
            The George Washington University, Washington, DC 20052, USA}

\author {Moskov~J.~Amaryan}
\affiliation{\ODU}
\author {William~J.~Briscoe}
\affiliation{\GWU}
\author {Michael~G.~Ryskin}
\affiliation{\PNPI}
\author {Igor~I.~Strakovsky}
\email[Corresponding author: ]{igor@gwu.edu}
\affiliation{\GWU}

\begin{abstract}
\vspace{0.5cm}
We evaluated CLAS Collaboration measurements for the $90^\circ$ meson photoproduction off the nucleon using a tagged photon beam spanning the energy interval $s = 3 - 11$~GeV$^2$. The results are compared with the "Quark Counting Rules" predictions.
\end{abstract}

\maketitle

\section{Introduction}
\label{sec:intro}

Binary reactions in QCD with large momentum transfer involve quark and gluon exchanges between colliding particles. The quark counting rule (QCR) of Brodsky and Farrar~\cite{Brodsky:1973kr} and Matveev, Muradyan, and Tavkhelidze~\cite{Matveev:1973ra} has a simple recipe to predict the energy dependence of the differential cross sections of two-body reactions at large meson production angles when $t/s$ is finite and is kept constant.  The fixed production or scattering angle behavior for exclusive processes is expected to be~\cite{Lepage:1980fj,Sivers:1974ux}
\begin{eqnarray}
        d\sigma/dt(s)\propto s^{-(n - 2)}~,
        \label{eq:eq1}
\end{eqnarray}
where $n$ is the minimum number of fundamental constituents (quarks) and $s$, $t$, and $u$ are Mandelstam variables.
After the Brodsky-Lepage~\cite{Lepage:1980fj}, the hard elastic scattering in QCD and the corrections to the leading behaviour were intensively discussed (see, e.g.,~\cite{Botts:1989kf,Botts:1989nd,Radyushkin:2009wx} and the references therein). If the photon is assumed to be one elementary field, then the prediction for a meson photoproduction is
\begin{eqnarray}
        d\sigma/dt(s)\propto s^{-7}~.
        \label{eq:eq2}
\end{eqnarray}
For the hadron-proton interaction, the counting rule works well, where hadron is a pion, kaon, proton, or antiproton~\cite{Sivers:1974ux,Jenkins:1979nc,Baglin:1982tu,White:1994tj,Armstrong:1997gv}.  The light meson photoproduction was examined in terms of the counting rule in Refs.~\cite{Anderson:1976ph,Jenkins:1995bk,Battaglieri:2001xv,Zhu:2002su,Battaglieri:2002pr,Chen:2009sda,Schumacher:2010qx,Kong:2015yzn,Dey:2014mka,Kunkel:2017src,Reed:2020dpf}. As has been observed, first of all at SLAC by Anderson \textit{et al.}, the reaction $\gamma p\to\pi^+n$ ($s = 8.4 - 15$~GeV$^2$) shows agreement with constituent counting rules that predict the cross section should vary as $s^{-7}$ and (n - 2) = $7.3\pm 0.4$~\cite{Anderson:1976ph}. The agreement extends down to $s$ = 6~GeV$^2$, where baryon resonances are still playing a role.\footnote{Potoproduction of K-mesons were considered in terms of QCR in Ref.~\cite{Chang:2015ioc}}.

Note, however, that the Quark Counting Rules account for the minimum numbers of elementary hard processes needed to provide a large momentum transfer to the hadron. At a very large energies, these rules are modified by the so-called Sudakov form factor~\cite{Dokshitzer:1991wu}.

Indeed, it is very improbable that two ensembles of constituents can get a strong transverse kick and radiate no gluons. Of course, the probability of a new gluon emission is suppressed by the QCD coupling constant $\alpha_s$, but simultaneously it can be enhanced by the square of the large logarithm - $\ln^2s$. The probability not to emit any additional gluons is called the Sudakov form factor. Thus for a very large $s$, we expect that the cross section of the large angle hadron-hadron scattering should fall down with $s$ faster than the QCR prediction~\cite{Botts:1989kf,Botts:1989nd}. The role of Sudakov form factor in large angle $pp$ elastic scattering was considered in Refs.~\cite{Ralston:1982pa,Pire:1982iv}.

On the other hand, it was shown in Ref.~\cite{Farrar:1989wb}  that due to the point-like nature of the photon, the Sudakov form factor is absent in the case of large angle photoproduction. Thus, photoproduction allows one to check the QCR directly in its original form.

In the present paper, we examined how the counting rules are applicable to the lightest meson photoproduction off the nucleon up to $s = 11$~GeV$^2$, where modern data are available mostly produced by the CLAS Collaboration at Jefferson Laboratory. The minimum value of $s$ in all these data exceeds 3~GeV$^2$.

Recall that there are three options of how one can consider a photon when it interacts to nucleon:
\begin{itemize}
\item No constituents ($n_\gamma = 0$) or $d\sigma/dt(s)\propto s^{-6}$,
\item Photon is a point-like particle which participate the strong interaction ($n_\gamma = 1$) or
      $d\sigma/dt(s)\propto s^{-7}$,
\item There is a $q-\bar{q}$ configuration which actually participates in the interaction         
      ($n_\gamma = 2$) or $d\sigma/dt(s)\propto s^{-8}$ 
\end{itemize}

\section{Light Meson Photoproduction Reactions}
\label{sec:data}

The JLab6 era has ended at Jefferson Laboratory leaving in its wake a plethora of cross section measurements for light meson photoproduction off the nucleon. There is a unique opportunity to bridge the resonance and high-energy regions, in particular that encompassing the region in which "Regge" theory is applicable,  and evaluate the quark counting rule phenomenology with differential cross sections above the "resonance" energies. 

\begin{figure}[htb!]
\centerline{
        \includegraphics[width=6in, angle=90]{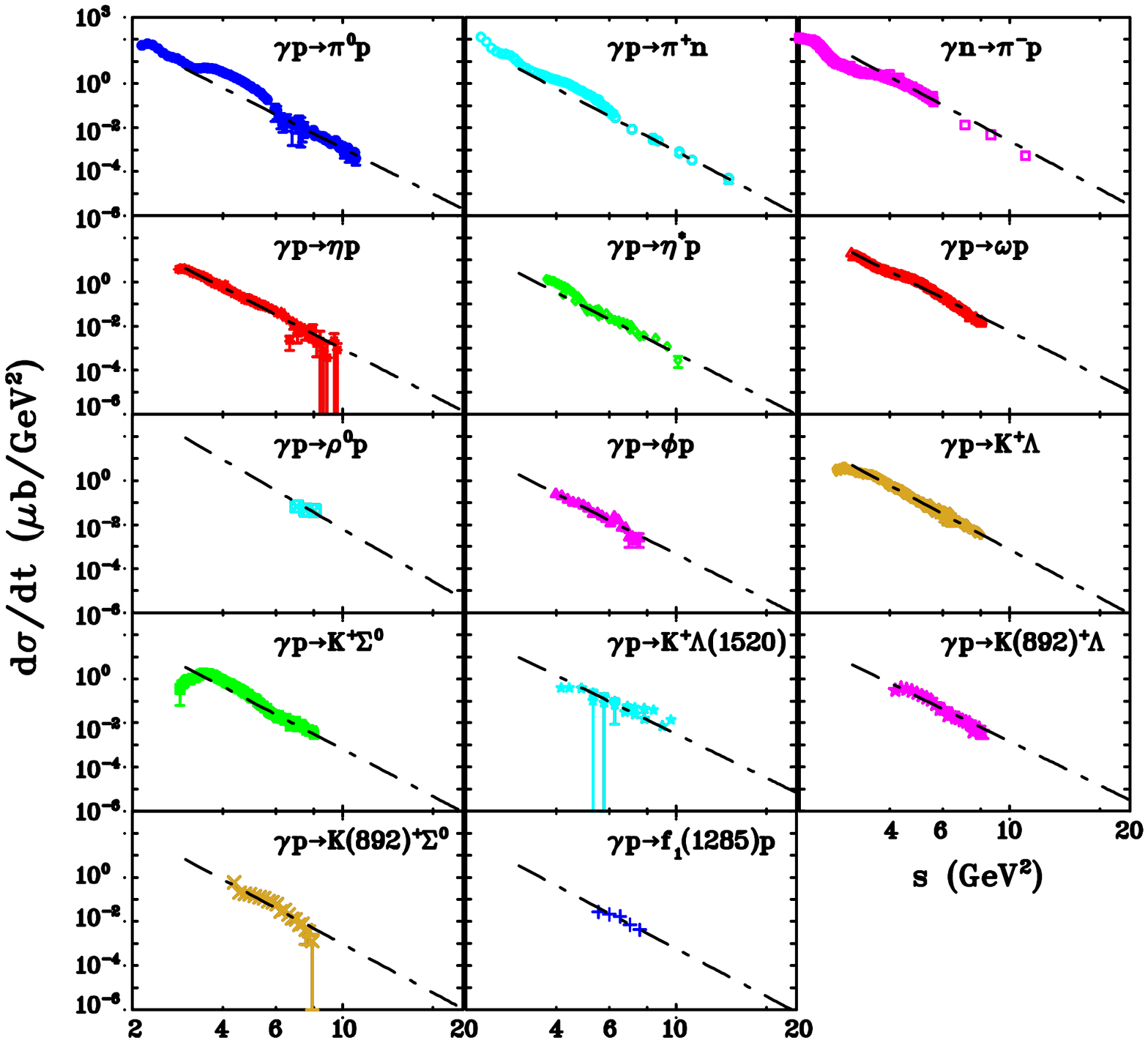}
}

\vspace{-1cm}
        \caption {Differential cross section of $\gamma N\to M B$, d$\sigma$/dt, at large 
                meson production angle $\theta = 90^\circ$ in c.m. as a function of invariant
                energy squared, $s$ (here $M$ is a meson and $B$ is a baryon). 
                Data are
$\gamma p\to\pi^0p$~\protect\cite{Kunkel:2017src,Dugger:2007bt}
	(blue filled circles),
$\gamma p\to\pi^+n$~\protect\cite{Dugger:2009pn,Zhu:2002su,Anderson:1976ph}
	(cyan open circles),
$\gamma p\to\pi^-p$~\protect\cite{Mattione:2017fxc,Zhu:2002su}
	(magenta open squares),
$\gamma p\to\eta p$~\protect\cite{Williams:2009yj,Hu:2020ecf}
	(red open asterisks),
$\gamma p\to\eta\prime p$~\protect\cite{Sowa:2016cs,Dickson:2016gwc,Williams:2009yj}
	(green open diamonds),
$\gamma p\to\omega p$~\protect\cite{Williams:2009ab,Battaglieri:2002pr}
	(red open triangles),
$\gamma p\to\rho^0 p$~\protect\cite{Battaglieri:2001xv},
    (cyan open squares with crosses),
$\gamma p\to\phi p$~\protect\cite{Dey:2014tfa}
	(magenta filed triangles), 
$\gamma p\to K^+\Lambda$~\protect\cite{McCracken:2009ra}
	(yellow filled diamonds),
$\gamma p\to K^+\Sigma^0$~\protect\cite{Dey:2010hh}
	(green filled squares), 
$\gamma p\to K^+\Lambda(1520)$~\protect\cite{Shrestha:2021hue,Moriya:2013hwg}
	(cyan open stars), 
$\gamma p\to K(892)^+\Lambda$~\protect\cite{Tang:2013gsa}
	(magenta filled stars),
$\gamma p\to K(892)^+\Sigma^0$~\protect\cite{Tang:2013gsa}
	(yellow crosses), and
$\gamma p\to\ f_1(1285)p$~\protect\cite{Dickson:2016gwc}
	(blue crosses).
	The black dash-dotted line is a result of the best-fits
	summarized in Table~\protect\ref{tab:tbl2}. 
	In the case of the $\omega$, the result corresponds to 
	the full energy range, $s = 3.5 - 8.1$~GeV$^2$.}
	\label{fig:g1}
\end{figure}

The new CLAS high statistical cross sections, for instance, obtained recently for $\gamma p\to\pi^0p$~\cite{Kunkel:2017src} are compared in Fig.~\ref{fig:g1} (top left) with previous data from CLAS measurements~\cite{Dugger:2007bt}.  At higher energies (above $s\sim6$~GeV$^2$) and large pion production angles ($\theta = 90^\circ$) in center-of-mass (c.m.) frame, the results are consistent with the $s^{-7}$ scaling expected from the QCR. The black dash-dotted line is a result of the best-fit of new CLAS data only~\cite{Kunkel:2017src}, performed with power function $\propto s^{-(n - 2)}$, leading to $(n - 2) = 6.89\pm 0.26$ (Table~\ref{tab:tbl2}).

The previous CLAS study for $\rho^0$~\cite{Battaglieri:2001xv} and $\omega$~\cite{Battaglieri:2002pr} results in $(n - 2) = 7.9\pm 0.3$ and $7.2\pm 0.7$, respectively. Mesons were identifies via the $\rho^0\to\pi^+\pi^-$ and $\omega\to\pi^+\pi^-\pi^0$ channels. \textit{N.B.} that the database for these analyses was limited by $s = 6.8 - 8.4$~GeV$^2$ and divided into 3-4 energy bins. Then the joint analysis of the CLAS~\cite{McCracken:2009ra} and SLAC~\cite{Anderson:1976ph} cross sections for for the reaction $\gamma p\to K^+\Lambda$ covering the range $s = 4.6 - 12.2$~GeV$^2$ gave $(n - 2) = 7.1\pm 0.1$~\cite{Schumacher:2010qx}. All these CLAS results are consistent with $s^{-7}$.
\begin{table}[th]
\caption{Power factor $(n - 2)$ in Eq.~(\ref{eq:eq1}) for light meson photoproduction
    off the nucleon came from the CLAS Collaboration. 
	The top (bottom) part summarizes pseudoscalar-mesons (vector-meson) results.    
    1st column gave reactions and 
    4th column shown best-fit results for the energy $s$ ranges listed in the
    2nd column and $|t|$ ranges given in the
    3rd column. Sources are given in the
	5th column. 
	To perform the best-fit for $\gamma p\to\pi^+n$, we added SLAC 
	data~\protect\cite{Anderson:1976ph} to JLab Hall~A
	measurements~\protect\cite{Zhu:2002su,Dugger:2009pn}.	
    As it is seen from Fig.~\protect\ref{fig:g1} in the case of a 
    vector-meson production at a larger $s$ the power $(n -2)$ increases. This can be 
    explained by the fact that in order to form the {\em vector}-meson, we need to 
    flip the $s$-channel helicity of the quark as it is mentioned in the text below.
	\label{tab:tbl2}}
\vspace{2mm}
\begin{tabular}{|c|c|c|c|c|}
\hline
Reaction                 & $s$      & $|t|$    & (n--2) & Ref.\\
                         &(GeV$^2$) &(GeV$^2$) &        & \\
\hline
$\gamma p\to\pi^0p$      &5.9--11.1& 2.1--4.7 & 6.89$\pm$0.26 &\protect\cite{Kunkel:2017src}\\
$\gamma p\to\pi^+n$      &6.3--14.9& 2.3--6.6 & 7.14$\pm$0.22 &\protect\cite{Dugger:2009pn,Zhu:2002su,Anderson:1976ph}\\
$\gamma n\to\pi^-p$      &4.0--11.3& 0.2--4.6 & 7.29$\pm$0.14 &\protect\cite{Mattione:2017fxc,Zhu:2002su}\\
$\gamma p\to\eta p$      &3.2-- 9.6& 0.6--3.8 & 7.02$\pm$0.16 &\protect\cite{Hu:2020ecf}\\
$\gamma p\to\eta\prime p$&4.2-- 9.3& 0.8--2.6 & 6.92$\pm$0.22 &\protect\cite{Sowa:2016cs,Dickson:2016gwc,Williams:2009yj}\\
$\gamma p\to K^+\Lambda$ &4.0-- 8.0& 0.3--2.9 & 7.28$\pm$0.06 &\protect\cite{McCracken:2009ra}\\
$\gamma p\to K^+\Sigma^0$&5.2-- 8.0& 0.3--2.8 & 7.12$\pm$0.21 &\protect\cite{Dey:2010hh}\\
$\gamma p\to K^+\Lambda(1520)$ &4.8-- 7.8& 0.9--3.2 & 6.65$\pm$0.41 &\protect\cite{Shrestha:2021hue,Moriya:2013hwg}\\
\hline
$\gamma p\to\omega p$    &3.5-- 8.1& 0.3--2.9 & 6.80$\pm$0.11 &\protect\cite{Williams:2009ab,Battaglieri:2002pr}\\
$\gamma p\to\omega p$    &5.0-- 8.1& 0.3--2.9 & 8.80$\pm$0.06\footnote{This result is performed for higher energy $s$ range.} &\protect\cite{Williams:2009ab,Battaglieri:2002pr}\\
$\gamma p\to\rho^0p$     &7.0-- 8.0& 2.3--2.9 & 7.9$\pm$0.3\footnote{This best-fit result taken from Ref.\protect\cite{Battaglieri:2001xv}.} &\protect\cite{Battaglieri:2001xv}\\
$\gamma p\to\phi p$      &4.0-- 7.5& 0.6--2.4 & 6.86$\pm$0.22 &\protect\cite{Dey:2014tfa}\\
$\gamma p\to K(892)^+\Lambda$  &4.2-- 8.1& 0.7--2.6 & 6.65$\pm$0.38&\protect\cite{Tang:2013gsa}\\
$\gamma p\to K(892)^+\Sigma^0$ &4.3-- 7.9& 0.7--2.4 & 7.34$\pm$0.45&\protect\cite{Tang:2013gsa}\\
$\gamma p\to f_1(1285)p$ &6.0-- 7.6& 1.2--2.0 & 7.19$\pm$0.96 &\protect\cite{Dickson:2016gwc}\\
\hline
\end{tabular}
\end{table}
\begin{figure}[htb!]
\centerline{
        \includegraphics[height=0.6\textwidth, angle=90]{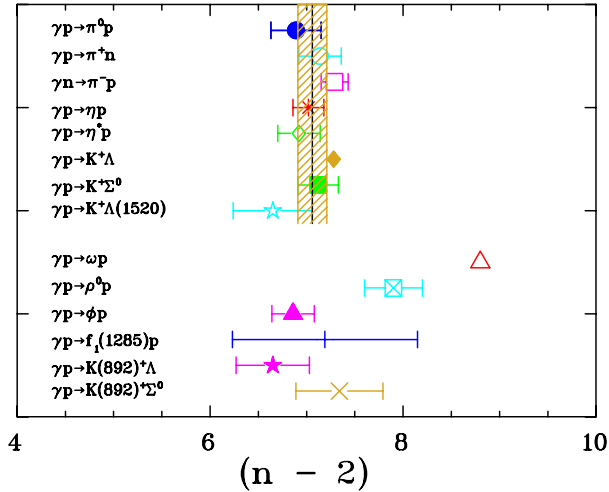}
}

\vspace{-0.5cm}
        \caption {Power factor $(n - 2)$ in Eq.~(\ref{eq:eq1}) for light meson 
        photoproduction off the nucleon came from the CLAS Collaboration.
        Black solid vertical line shows average value for pseudoscalar-mesons
        $<(n - 2)> = 7.06\pm 0.15$.
        Yellow band presents its uncertainty. In the case of the 
        $\omega$, the result corresponds to the higher energy range, $s = 5 - 8.1$~GeV$^2$.
		The notation for the different reactions is the same as in 
		Fig.~\protect\ref{fig:g1}.} \label{fig:g1a}
\end{figure}

For our analysis, we included a number of light meson photoproduction data sets produced by the CLAS Collaboration on the proton and neutron for incident photon energies above $s = 3$~GeV$^2$, carried out during the past 20 years. For one particular case, the $\gamma p\to\pi^+n$ analysis, we included JLab Hall~A~\cite{Zhu:2002su} and SLAC~\cite{Anderson:1976ph} measurements. The results (Fig.~\ref{fig:g1} and Table~\ref{tab:tbl2}) are consistent with the $(n - 2) = 7$ (see Fig.~\ref{fig:g1a}) scaling expected from the QCR. Oscillations observed at low energies indicate that the QCR requires higher energies and higher $|t|$ and $|u|$ before it can provide a valid description. Obviously, the extended energy range would be more definitive; our results do appear to be consistent with this limit. The JLab12 and EIC programs are capable of providing the data needed to improve our results.

Recently, the analysis of the CLAS $\gamma p\to\eta\prime p$, $\gamma p\to K^+\Lambda$, and $\gamma p\to K^+\Sigma^0$~\cite{Dey:2014mka} data covered a limited energy range of $s = 6.2 - 8.1$~GeV$^2$ shown that the energy behaviour of $90^\circ$ cross section is $d\sigma/dt(s)\propto s^{-7}$. While in the case of $\gamma p\to\eta p$, $\gamma p\to\omega p$, and $\gamma p\to\phi p$, results are $(n - 2) = 12.7\pm 1.2$, $(n - 2) = 9.4\pm 0.1$, and $(n - 2) = 12.3\pm 0.6$, respectively. Mesons were identified via $\eta\to\pi^+\pi^-\pi^0$, $\omega\to\pi^+\pi^-\pi^0$, and $\phi\to K^+K^-$, respectively.   Then the analysis of the $\omega$ photoproduction data~\cite{Williams:2009ab} for $s = 5 - 8$~GeV$^2$ results $(n - 2) = 9.08\pm 0.11$~\cite{Reed:2020dpf}.

The A2 Collaboration at MAMI reported differential cross sections for the $\gamma p\to\omega p$ close to threshold~\cite{Strakovsky:2014wja}.  The omega-meson was identified via a radiative decay mode $\omega\to\pi^0\gamma$. As Figure~4 of Ref.~\cite{Strakovsky:2014wja} shows, there is a disagreement between CLAS and A2 measurements below $s = 3$~GeV$^2$.

\begin{figure}[htb!]
   \label{f4}
        \includegraphics[scale=0.2]{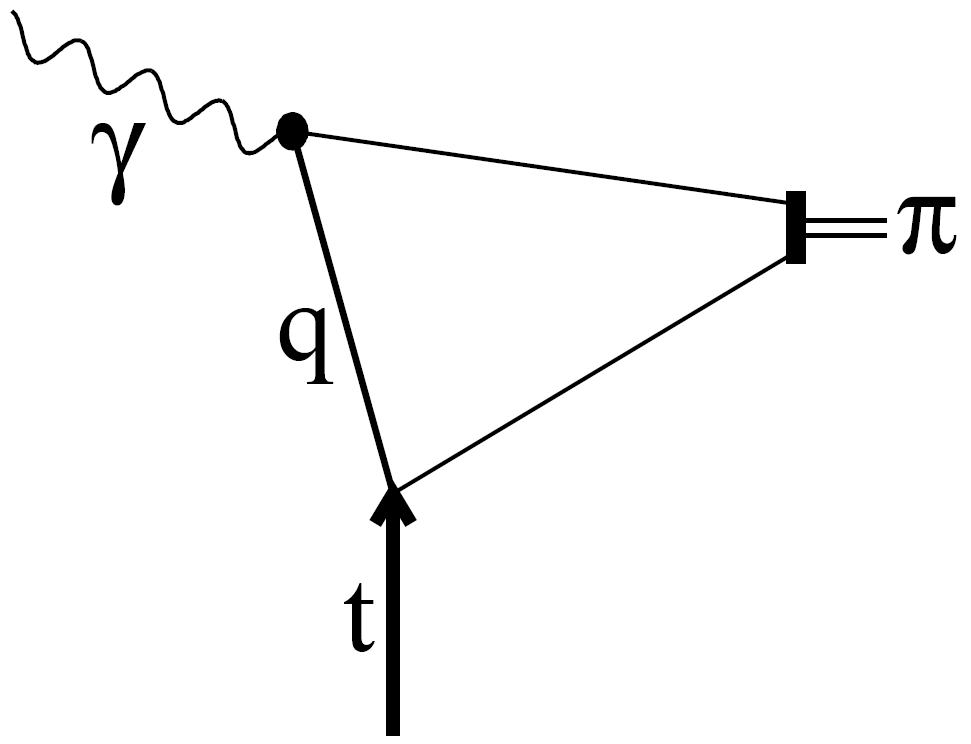}

        \caption {Simplest diagram for the large $t$ meson photoproduction.}
        \label{fig:g4}
\end{figure}

The difference between our analysis and analysis by Dey~\cite{Dey:2014mka} who obtained a larger power $(n - 2)$ for the reactions $\gamma p\to\eta p$, $\gamma p\to\omega p$, and $\gamma p\to\phi p$ is understandable due to different energy ranges of the data included in the  fits. Indeed, as one  can  see in Fig.~\ref{fig:g1} for these reactions, there is a steeper energy dependence of the higher $s$ part of the distribution. For the case of the $\phi$ (and partly $\eta$) photoproduction, this can be considered as a hint in favour of the noticeable role of the five quark ($uuds\bar s$) component in the proton wave function. Having such a component, the process can be considered as the constituent strange quark interchange between the proton and the $\phi$ meson. However, this explanation is not directly applicable to the $\omega$ photoproduction. 
Note, however, that, as was mentioned in Ref.~\cite{Reed:2020dpf}, due to a vector nature of $\omega$ meson
in order to form the spin part of the corresponding wave function, we have to violate the $s$-channel helicity conservation (SCHC). Therefore, we have to expect an additional suppression of  90$^\circ$ high energy photoproduction~\cite{Lepage:1980fj}. That is for the case of the $\rho$, $\omega$, and $\phi$ mesons the expected power $(n-2)$ should be 1 (for the case of Fig.~\ref{fig:g4} diagram) or  2 (for the case of Fig.~8 of Ref.~\cite{Reed:2020dpf}) units larger). Thus we can say that the observed energy dependence of $\rho$ cross section and the behaviour of $\omega$ at a larger $s$ is consistent with the QCR.

Recall that in the case of $\gamma p\to\omega p$ and $\gamma p\to\phi p$, both analyses (our and \cite{Dey:2014mka}) used the same experimental data.  This indicates the  necessity of more experimental data in a wider energy range, especially for the $\rho$ and $\omega$ mesons photoproduction, to better study the energy dependence, the role of the meson spin and to obtain a more stringent constraint on the fit parameters.  It will be interesting also to measure and compare the angular behaviour to study the elementary quark-quark hard collision. 

In particular, it is important to compare the $\rho$ and $\omega$ vector-mesons photoproduction in a larger energy interval. First, the $s = 7 - 8$~GeV$^2$ interval, where the $\rho$  cross section was measured up to now~\cite{Battaglieri:2001xv}, is not sufficient to reliably determine the power $(n - 2)$. Note also that fitting the $\omega$ data at a larger $s = 5 - 8$~GeV$^2$, we get a larger value of $(n - 2) = 8.80\pm 0.06$ which is consistent with results of Refs.~\cite{Dey:2014mka,Reed:2020dpf} and the QCR prediction ($(n - 2) = 9$) for the process with SCHC violation.

\begin{figure}[htb!]
\centerline{
        \includegraphics[height=0.9\textwidth, angle=90]{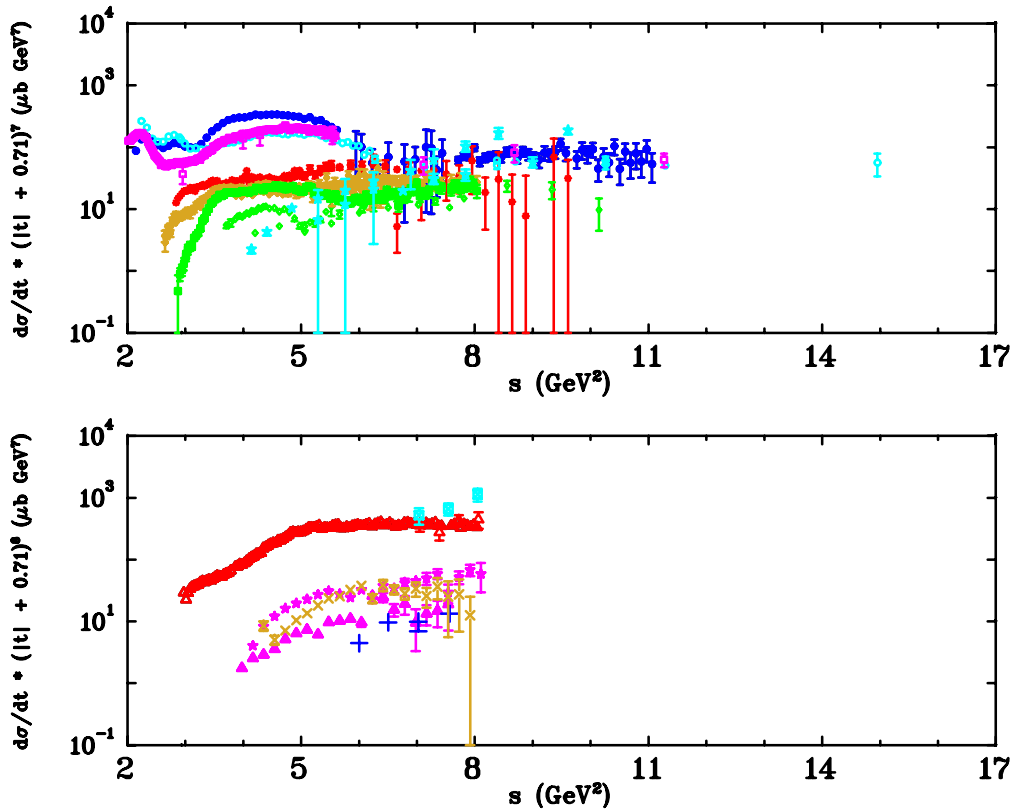}
}

\vspace{-0.5cm}
        \caption {Differential cross section of the light meson production 
		off the nucleon at meson production angle $\theta = 90^\circ$ in c.m. 
		as a function of c.m. energy squared $s$. 
		\underline{Upper Panel}: Pseudoscalar-mesons d$\sigma/dt\cdot (t - 0.71)^7$
		and 
		\underline{Lower Panel}: Vector-mesons d$\sigma/dt\cdot (t - 0.71)^8$. 
		The notation for the different reactions is the same as in 
		Fig.~\protect\ref{fig:g1}. } \label{fig:g2}
\end{figure}

Next the ratio of $\sigma(\rho)/\sigma(\omega)$  cross sections at the moment looks a bit strange. If the photon first produces the $q\bar q$ pair then we expect $\sigma(\rho)/\sigma(\omega) = 9$, like in diffractive photoproduction. If on contrary the photon interacts with one of the proton's valence quarks then the ratio  $\sigma(\rho)/\sigma(\omega) = (5/3)^2$. However, as it is seen in Fig.~\ref{fig:g2} at $s = 7$ GeV$^2$, we observe $\sigma(\rho)\simeq \sigma(\omega)$.\\

The additional interesting fact is that the $\phi$, $f_1(1285)$, and $K(892)^+$ cross sections are close to each other (Fig.~\ref{fig:g2}).\\

Since we consider not very large $s$, we have to discuss the possible power corrections to the QCR. Unfortunately, the corresponding power corrections are closely related to the non perturbative structure of incoming hadrons. Therefore, we evaluate the possible role/scale of the power corrections based on the well known dipole behaviour of the proton QED form factor $G(t) = (1 - t/0.71\mbox{GeV}^2)^2$, which describes all four-momentum dependencies of both electric and magnetic form factors of proton quite well~\cite{Halzen:1984mc}, where the constant 0.71~GeV$^2$ determines the scale of the correction in comparison with the asymptotic behaviour $G(t) = 1/t^2$. It appears natural to introduce a similar ``infrared cutoff'' at a lower $s$ in our case as well.
Thus in Fig.~\ref{fig:g2} (upper panel), we plot the product $d\sigma/dt(s)\cdot (t - 0.71)^7$ for the pseudoscalar-mesons. As it is seen, now the $s$ behaviour of this product is rather flat down to $s= 2 - 3$~GeV$^2$.
In the lower panel for the vector~mesons, we plot the product $d\sigma/dt(s)\cdot (t - 0.71)^8$. Here we see the flat $\omega$ cross section dependence for $s > 5$~GeV$^2$.  Additionally, Figure~\ref{fig:g2} shows that the accuracy (and dispersion) of the data points is better seen here than on Fig.~\ref{fig:g1} and it demonstrates the possible role of the "infrared cutoff" $(t - 0.71)$ in this energy interval.

\begin{figure}[htb!]
\centerline{
        \includegraphics[height=0.6\textwidth, angle=90]{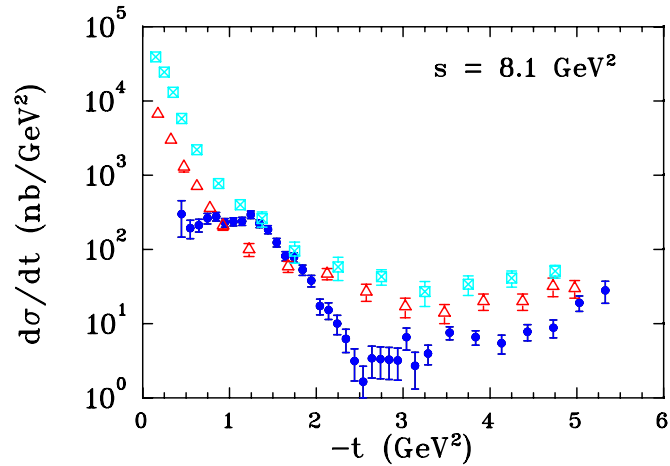}
}

\vspace{-0.7cm}
        \caption {Differential cross section at $s = 8.1$~GeV$^2$ 
        for the reaction $\gamma p\to\pi^0p$ for $|t| = 0.5 - 5.3$~GeV$^2$ ($90^\circ$
        corresponds to $|t| = 3.2$~GeV$^2$) shown by the blue filled circles~\protect\cite{Kunkel:2017src}, for the reaction $\gamma p\to\omega p$ for $|t| = 0.2 - 5.0$~GeV$^2$ ($90^\circ$ corresponds to $|t| = 3.0$~GeV$^2$) shown by red open triangles~\protect\cite{Battaglieri:2002pr}, and
        for the reaction $\gamma p\to\rho p$ for $|t| = 0.2 - 4.7$~GeV$^2$ ($90^\circ$ corresponds to $|t| = 3.0$~GeV$^2$) shown by cyan open squares with crosses~\protect\cite{Battaglieri:2001xv}. } \label{fig:g3}
\end{figure}

In Figure~\ref{fig:g3}, the differential cross section $d\sigma/dt$ is plotted as a function of the $t$-Mandelstam at $s = 8.1$~GeV$^2$. The red open triangles are measured data points of the $\gamma p\to\omega p$ reaction~\cite{Battaglieri:2002pr} and cyan open squares with crosses of the $\gamma p\to\rho p$ reaction~\cite{Battaglieri:2001xv}, while the blue solid circles are $\pi^0$ photoproduction  data points~\cite{Kunkel:2017src}. As one can see for the pseudoscalar-meson production at lower values of $t$, the cross section of $\omega$ and $\rho$ photoproduction is an order of magnitude higher than that of $\pi^0$ photoproduction, however at higher values of $t$, the $\omega$ photoproduction cross section is in general still higher but the difference is not as dramatic as at lower values of $t$. It has to be mentioned that data on this figure are for all meson production angles. When only the $90^{\circ}$ production angle data are selected, these three cross sections at higher values of $s$ reach the same level as well as all other meson production data, except of the $\phi$ and the $f_1(1285)$ photoprodution cross sections, which lie significantly below the other mesons plateau at higher energies (Fig.~\ref{fig:g2}).

It is interesting to see that the $\phi$ and $f_1(1285)$ production cross sections at higher energies and the $90^{\circ}$ production angle are equal to each other within statistical errors, which may indicate a common mechanism of their production. While the $\phi$ has definitely $s\bar s$ quark structure, the $K\bar K \pi$ branching ratio of $f_1(1285)$ is on the order of 10$\%$ which means that in average the $s\bar s$ component of the wave function of the $f_1(1285)$ is small. However, it seems that, as presented data may indicate, the $s\bar s$ component of the $f_1(1285)$ wave function becomes dominant in the hard scattering process, when all three Mandelstam variables $s$, $t$, and $u$ are large.

\section{Summary and Conclusions}
\label{sec:conc}

In the present paper, we study the energy dependence of the $90^\circ$ light meson photoproduction off the nucleon. We consider practically all available experimental data obtained by the CLAS Collaboration over more than the last two decades and compare the results with the Quark Counting Rules predictions. We emphasize that in the case of \textit{photoproduction} the QCR prediction does not affected by the Sudakov form factor. This fact allows a more direct interpretation of the observed results. 

Thanks to the point-like nature of the photon, the $90^\circ$ cross section $d\sigma/dt\propto s^{-7}$ for the pseudoscalar-meson production. The average value of $<(n-2)>$ for pseudoscalar-meson reactions listed in Table~\ref{tab:tbl2} (top) is $7.06\pm 0.15$. The explanation of the $s^{-7}$ instead of $s^{-8}$ or $s^{-6}$ is: 
\begin{itemize}
\item  In terms of Brodsky-Farrar~\cite{Brodsky:1973kr}: "In photoproduction amplitude, the 
    balance between the quarks momenta was provided by the highly virtual quark with propagator $1/\hat q\propto 1/\sqrt{s}$ (Fig.~\ref{fig:g4}) instead of the gluon for which the propagator is $\propto 1/s$."
\item It terms of Matveev \textit{et al.}~\cite{Matveev:1973ra}: "In photoproduction, the incoming
    $q\bar{q}$ pair is produced (in the case of a large momentum transferred) very close to the interaction point and not in advance (at a large distance) as in the vector-meson dominance model. That is in the incoming state, we deal with a "point-like" $q\bar{q}$ pair and only in the final state we have two quarks separated by a large ($\sim 0.5 - 1$~fm) distances. The small factor corresponding to the probability to have two quarks very close to each other is needed now for the final state only (and not for the initial state). This leads to the root square of the usual $1/s^2$ factor."
\end{itemize}

Let us note that the cross sections for the light meson photoproduction off the nucleon at 90$^\circ$ is very small (minimal) and for that reason it may cause a problem for the best-fit analysis using Eq.~(\ref{eq:eq2}).

Obviously, the JLab6 program is limited by $s\leq~11$~GeV$^2$. Within the JLab12 program, the approved by JLab PAC proposal E12–-14–-005 for Hall~C can extend the measurement of the $\gamma p\to\pi^0p$ reaction up to $s\leq~20$~GeV$^2$~\cite{Dutta:2014dd}. Additionally, recent GlueX Collaboration reports allow expect data for $\gamma p\to\eta p$~\cite{Kamel:2021} and $\gamma p\to\omega p$~\cite{Dalton:2021}.

\acknowledgments

We thank Stan Brodsky, Boris Kopeliovich, Shunzo Kumano, Fred Myhrer, and Bernard Pire for useful remarks and discussions.  This work was supported in part by the by the U.~S.~Department of Energy, Office of Science, Office of Nuclear Physics, under Award Numbers DE--SC0016583 and DE--FG02--96ER40960.



\end{document}